\RequirePackage{ifpdf}
\ifpdf 
\documentclass[pdftex]{sigma}
\else
\documentclass{sigma}
\fi

\begin{document}

\renewcommand{\thefootnote}{$*$}

\renewcommand{\PaperNumber}{054}

\FirstPageHeading

\ShortArticleName{Wigner Function and Canonical Transformations in Time-Dependent Quantum Mechanics}

\ArticleName{Wigner Distribution Functions\\ and the Representation of Canonical Transformations\\ in Time-Dependent Quantum Mechanics}

\Author{Dieter SCHUCH~$^{\dag\ddag}$ and Marcos MOSHINSKY~$^\ddag$\footnote{Member of El Colegio Nacional and Sistema Nacional de Investigadores.}}

\AuthorNameForHeading{D. Schuch and M. Moshinsky}

\Address{$^\dag$~Institut f\"ur
Theoretische Physik, Goethe-Universit\"at Frankfurt am Main,\\
$\phantom{^\dag}$~Max-von-Laue-Str.~1, D-60438 Frankfurt am Main, Germany}
\EmailD{\href{mailto:schuch@em.uni-frankfurt.de}{schuch@em.uni-frankfurt.de}}

\Address{$^\ddag$~Instituto de F\'{\i}sica, Universidad Nacional Aut\'onoma de
  M\'exico,\\
$\phantom{^\ddag}$~Apartado Postal  20-364, 01000 M\'exico D.F., M\'exico}

\EmailD{\href{mailto:moshi@fisica.unam.mx}{moshi@fisica.unam.mx}}

\ArticleDates{Received February 06, 2008, in f\/inal form June 08,
2008; Published online July 14, 2008}

\Abstract{For classical canonical
transformations, one can, using the Wigner transformation, pass
from their representation in Hilbert space to a kernel in phase
space. In this paper it will be discussed how the time-dependence of the
uncertainties of the corresponding time-dependent quantum problems can
be incorporated into this formalism.}


\Classification{37J15; 81Q05; 81R05; 81S30}

\vspace{-2mm}

\section{Introduction}

In classical Hamiltonian mechanics the time-evolution of a physical system is
described by canonical transformations in phase space that keep the Poisson
brackets of the transformed coordinate and momentum with respect to the
initial ones unchanged. This transformation in phase space can be described
(for a one-dimensional problem in physical space and, therefore, a
two-dimensional one in phase space, to which we will restrict ourselves in the
following) by the so-called two-dimensional real symplectic group $Sp(2,{\mathbb R})$,
represented by  $2 \times 2$ matrices with determinant equal to 1. (In order
to compare our new results with earlier ones for the time-independent case, we
only consider the homogeneous symplectic group without translations, not the
inhomogeneous symplectic group $ISp(2,{\mathbb R})$.) It has been shown in \cite{1} how
it is possible to obtain the representation of the group of linear canonical
transformations in time-independent quantum mechanics via the determination of
the conf\/iguration space representation of the unitary operator that connects
quantum mechanically the transformed variables $x$ and $p$ with the initial
ones, $x'$ and $p'$. A subsequent Wigner transformation shows explicitly that
for the time-independent problems considered by this method, essentially the
classical results are reproduced. This agrees with the fact that at least
for quadratic Hamiltonians, the Wigner function evolves as $W(x',p',t) =
W(x_M(x',p',-t),p_M(x',p',-t),0)$, where
$x_M$ and $p_M$ are the Moyal time evolution of position and momentum which,
again for quadratic Hamiltonians, coincide with the classical evolution \cite{2,3,4,5}.

In a dif\/ferent study \cite{6} it has been shown that, for time-dependent
quantum systems, cha\-racteristic dif\/ferences compared with the classical
situation arise, in particular, when the time-dependence of the quantum
mechanical uncertainties of position and momentum are taken into account. One
such dif\/ference can already be found for the most simple physical system, the
free motion.

In classical mechanics, the dynamics of a system may not only be
described by its trajectory but, particularly for systems with many degrees
of freedom, a statistical description in terms of virtual ensembles or
distribution functions in phase space (also called $\Gamma$-space in this
context) is possible. These density-type functions $\varrho_{\Gamma}(x,p,t)$ must be globally
conserved, which is guaranteed if they fulf\/il a kind of conservation law in the
form of a continuity equation in phase space that connects the explicit
change in time of the density function $\varrho_{\Gamma}$ with the divergence of a current
$\vec{j}_{\Gamma} = \varrho_{\Gamma} \vec{v}_{\Gamma}$. If the system obeys Hamilton's equations of motion, the divergence of
the velocity f\/ield $\vec{v}_{\Gamma}$ always vanishes, $\nabla_{\Gamma}  \vec{v}_{\Gamma}  = 0$ (where $\nabla_{\Gamma}$ is the
nabla-operator in phase space). In a~hydrodynamical description, this would
mean that $\varrho_{\Gamma}$ describes an incompressible medium. In this probabilistic
context, this corresponds to Liouville's theorem that, from all
possible transformations of phase space, selects only those where a phase extension
always retains its volume during motion. Making use of the concepts of measure
theory, this statement can be made even more precise in the formulation that
the measure of point-sets is an invariant of the time-evolution of the virtual
ensemble.

It can be shown straightforwardly that the quantum mechanical density $\varrho
(x{,}t) {=} \Psi^{\ast} (x{,}t) \Psi (x{,}t)$,
corresponding to the complex solution $\Psi (x,t)$ of the time-dependent
Schr\"odinger equation also fulf\/ils a continuity equation (now in position-
or conf\/iguration space). In this case, however, the divergence of the
corresponding vector f\/ield is proportional to the relative change in time of
the mean square deviation of position (position uncertainty) $\langle \tilde{x}^2\rangle = \langle x^2\rangle  - \langle x\rangle^2$, i.e.,
$\nabla_x \vec{v} = \frac{1}{2} \frac{d}{dt}\langle \tilde{x}^2\rangle/ \langle \tilde{x}^2\rangle$. So,
$\nabla_x \vec{v}$ only vanishes if $\langle \tilde{x}^2\rangle$ is constant which, e.g.\ for a
Gaussian wave packet, would correspond to a constant width of this wave
packet. It is, however, well known that already for the free motion, the wave
packet width is not constant but spreading with increasing time. If one
would consider the wave packet width as a kind of measure for the ``volume''
of our quantum system, it certainly would not be conserved under
time-evolution. This situation also does not change if one tries to include
the momentum aspect by also taking into account the corresponding mean-square
deviation $\langle\tilde{p}^2\rangle$ and then considering the product of $\langle \tilde{x}^2\rangle $ and $\langle \tilde{p}^2\rangle$ as an
adequate measure for the ``volume'' of the system; since $\langle\tilde{p}^2\rangle$ is constant for
the free motion, therefore, the product with $\langle \tilde{x}^2\rangle$ still grows in time. (For a
more detailed discussion, see also~\cite{6}.) A more consistent approach for a
comparison with the classical case could, however, make use of the phase-space
formulation of quantum mechanics in the form of the time-dependent Wigner
function which shall be done in this paper.

Unlike in the time-independent situation where the quantum mechanical results
mainly reproduce the classical ones, in the time-dependent case there are
obviously, at least, formal dif\/ferences between the classical and the quantum
mechanical descriptions of the system even already for such simple ones like the
free motion. These dif\/ferences are intimately connected with the
time-dependence of the typical quantum mechanical aspect of the system, namely,
the uncertainties of position and momentum. Therefore, in this paper,  we will investigate the inf\/luence of the
time-dependence of the quantum system, in particular of the uncertainties, on
the representation of the group of linear canonical transformations in quantum
mechanics.

For this purpose, in Section~\ref{sec2}, we brief\/ly summarize the main results of the
time-independent quantum mechanical case. In Section~\ref{sec3} we then consider the
time-dependent case and discuss the characteristic dif\/ferences compared with
the time-independent situation, in particular the role of the time-dependence
of the quantum uncertainties. In order to be on the safe side of systems
with exact analytic solutions, we restrict our discussion to systems with at
most quadratic Hamiltonians, in particular, to the harmonic oscillator (with
possibly time-dependent frequency) and the free motion. Possible ways of
overcoming these limitations and of extending our method to further problems will
be mentioned in Section~\ref{sec4} where the results will also be summarized and some
perspectives will be mentioned.

\section{Time-independent case}\label{sec2}

The time-evolution in classical Hamiltonian mechanics is
described by canonical transformations in phase space that can be represented by
\begin{gather}
\left(\begin{array}{c} x \\ p \end{array}\right)  = \left(\begin{array}{cc} a & b \\ c &
     d \end{array}\right)   \left(\begin{array}{c} x' \\ p' \end{array}\right),
\label{sm01}
\end{gather}
where $a$, $b$, $c$, $d$ are real and the determinant of the $2 \times 2$ matrix is $1$, i.e.,
$ad-bc=1$. The group of transformations represented by the $2 \times 2$ matrices is the
so-called two-dimensional real symplectic group $Sp(2,{\mathbb R})$.

Following \cite{1} (see Chapter~35 and references cited
therein), it is possible to obtain the representation of the group of
linear canonical transformations \eqref{sm01} in quantum mechanics. Referring to
\cite{1,7}, the main
objective is to determine the conf\/iguration space representation
\begin{gather*}
\langle x|U|x'\rangle = K(x,x')
\end{gather*}
of the unitary operator $U$ that provides the quantum mechanical relation
between $x$, $p$ and~$x'$,~$p'$, according to
\begin{gather}
x = U x' U^{-1},
\qquad
p = U p' U^{-1} .
\end{gather}

With the help of the kernel $K(x,x')$, the ef\/fect of any canonical
transformation \eqref{sm01} can be described as
\begin{gather*}
\Psi (x) = \int^{+\infty}_{-\infty} dx'\,
   K(x,x')
        \Psi (x') .
\end{gather*}

The integral kernel $K(x, x')$ has been derived taking into consideration the fact that it must
satisfy the following two dif\/ferential equations \cite{8,9}
\begin{gather}
\left(a x + b \frac{\hbar}{i} \frac{\partial}{\partial x} \right)
K(x,x') = x' K(x,x'),
\label{sm06}
\\
\left(c x + d \frac{\hbar}{i} \frac{\partial}{\partial x} \right)
K(x,x') = -\frac{\hbar}{i} \frac{\partial}{\partial x'} K(x,x') .
\label{sm07}
\end{gather}

An exponential ansatz, bilinear in $x$ and $x'$, f\/inally leads to $K(x,x')$ in
the form
\begin{gather}
 K (x, x') = \left(\frac{1}{2\pi b}\right)^{\frac{1}{2}} \exp \left\{
 - \frac{i}{2 b} \big[ a x^2 - 2 x x' + d x'^2 \big]\right\} .
\label{sm08}
\end{gather}

This kernel $K(x,x')$, related with the specif\/ic canonical transformation, is
formulated in conf\/iguration space, whereas, the corresponding classical
canonical transformation is formulated in phase space. Therefore, it is
interesting to discuss the representation of this canonical transformation in
the phase space version of quantum mechanics that was developed by Wigner~\cite{10} with the help of the corresponding distribution function. This
distribution function $W(x,p)$ can be obtained from a given wave function
$\Psi(x)$ in conf\/iguration space via the so-called Wigner transformation
\begin{gather}
W(x,p) = \frac{1}{2 \pi \hbar} \int^{+\infty}_{-\infty} dy\, e^{ipy / \hbar}
 \Psi^{\ast}
\left(x+\frac{y}{2}\right) \Psi \left(x-\frac{y}{2}\right) .
\label{sm09}
\end{gather}

Applying this transformation to the kernel \eqref{sm08} leads to the phase space kernel
in the form \cite{7}
\begin{gather*}
 K (x, x',p,p') = \delta [x' -(a x + b p)] \delta [p' - (c x + d p)],
\end{gather*}
showing that, for this linear canonical transformation, the kernel coincides
with its classical limit. So, the quantum mechanical problem mainly reproduces
the classical situation without any additional specif\/ic quantum mechanical
aspect.

\section{Time-dependent case}\label{sec3}

Now we will investigate how far this is  still true in the case of
specif\/ic quantum dynamical aspects entering the problem. So, considering
time-dependent problems in quantum mechanics in terms of the time-dependent
Schr\"{o}dinger equation or equivalent formulations, one f\/inds that not only
classical position and momentum change in time (in a way that can be described
by canonical transformations) but, also the typical quantum mechanical degrees
of freedom, like position- and momentum-uncertainties, may be time-dependent
(corresponding, e.g., to wave packets with time-dependent width). For certain
problems with exact analytical solutions in form of Gaussian wave packets, it
has been shown (see, e.g., \cite{11} and references cited
therein) how the transition from initial position and time (in conf\/iguration
space) to any later position and time can be achieved with the help of a
time-dependent kernel (or propagator) $K(x,x',t,t')$ according to
\begin{gather}
\Psi (x,t) = \int^{+\infty}_{-\infty} dx'\,
   K(x,x',t,t')
        \Psi (x',t').
\label{sm11}
\end{gather}

The integral kernel $K(x, x', t, t')$ can be obtained in dif\/ferent ways, e.g., via
Feynman's path integral method \cite{12}, or, for kernels quadratic in $x$
and $x'$, similar to the discussion in the time-independent case \cite{7} in the form
\begin{gather}
 K(x, x', t, t'=0) = \left(\frac{m}{2\pi i \hbar \alpha_0
      \hat z}\right)^{\frac{1}{2}} \exp \left\{ \frac{im}{2\hbar \hat z} \left[
      \dot{\hat z} x^2 - 2 x \left(
      \frac{x'}{\alpha_0}\right) + \hat u \left(
      \frac{x'}{\alpha_0}\right)^2 \right]\right\},
\label{sm12}
\end{gather}
where $\alpha_0 = \sqrt{\frac{2 m \langle \tilde{x}^2\rangle _0}{\hbar}}$ is proportional to the initial position uncertainty,
or initial mean square deviation in space, $\langle \tilde{x}^2\rangle_0 = \langle x^2\rangle_0 -
\langle x\rangle ^2_0$ (where $\langle \cdots\rangle$ denotes mean
values calculated according to $\langle\cdots\rangle = \int^{+\infty}_{-\infty}  \Psi^{\ast}
  \cdots \Psi\, dx'$). We choose this form of the kernel because of the explicit
  appearance of $\alpha_0$ and, therefore, the initial position uncertainty. This quantity is essential for the time-evolution of the
  quantum uncertainties. As has been shown in~\cite{11}, already for the simple
  example of the harmonic oscillator, the time-evolution of the position
  uncertainty and, hence, the width of the Gaussian wave packet solution of
  this problem, behaves qualitatively very dif\/ferent if the initial position
  uncertainty corresponds to that of the ground state (constant width) or
  dif\/fers from it (oscillating width) where the latter represents the
  general solution which, in the limit $\omega \rightarrow 0$, provides the
  free motion wave packet, whereas, the former only leads to a plane wave
  solution in this limit (for further details see~\cite{11}). The
  time-dependence enters this kernel explicitly
via the parameters $\hat{z}(t)$ and $\hat{u}(t)$. In the limit $t \rightarrow
0$, the kernel turns into a delta function.

Since, according to~\eqref{sm11}, the dependence of $\Psi(x,t)$ on $x$ and $t$ enters only
via $K(x,x',t,t')$, this kernel also must fulf\/il the time-dependent Schr\"{o}dinger
equation. Inserting $K$, in the form given in~\eqref{sm12}, into this equation shows that the parameters $\hat{z}(t)$
and $\hat{u}(t)$ both fulf\/il the equation of motion for the corresponding classical
problem (e.g., the free motion or the harmonic oscillator with possibly
time-dependent frequency), however, they are not independent of each other but
are coupled via the relation
\begin{gather}
\dot{\hat z} \hat u - \dot{\hat u} \hat z = 1.
\label{sm13}
\end{gather}

This resembles the condition that the determinant of the entries of the $2
\times 2$ matrix of the linear canonical transformation \eqref{sm01} must fulf\/il.

Inserting \eqref{sm12} into \eqref{sm11} with an initial Gaussian wave packet $\Psi_{\rm WP} (x',t')$ of
the form
\begin{gather*}
  \Psi_{\rm WP} (x',t'=0) = \left(\frac{m\beta_0}{\pi
            \hbar}\right)^\frac14 \exp \left\{ \frac{im}{2\hbar}
            \left[ i \left(\frac{x'}{\alpha_0}\right)^2+ 2 \frac{p_0}{m} x'
            \right]\right\}
\end{gather*}
leads to a
Gaussian wave packet
\begin{gather}
 \Psi_{\rm WP} (x,t) = \left(\frac{m}{\pi\hbar}\right)^\frac14
        \left(\frac{1}{\hat u + i \hat z}\right)^{\frac{1}{2}} \exp
        \left\{ \frac{im}{2\hbar} \left[ \frac{\dot{\hat z}}{\hat z}
        x^2 - \frac{(x-\frac{p_0\alpha_0}{m} \hat z)^2}{\hat z
        (\hat u + i \hat z)}\right]\right\},
\label{sm15}
\end{gather}
whose maximum follows the classical trajectory $\eta (t)$, which is, therefore, up to a
constant, proportional to $\hat{z}$, i.e.,
\begin{gather}
 \hat{z} = \frac{m}{\alpha_0 p_0} \langle x\rangle (t) = \frac{m}{\alpha_0 p_0} \eta (t)
\label{sm16}
\end{gather}
with initial momentum $p_0$ and mean value of position $\langle x\rangle$  which is, according
to Ehrenfest's theorem, identical with the classical trajectory $\eta (t)$.

The coef\/f\/icient of the term quadratic in $x$ in the exponent of the Gaussian
can be expressed via the complex quantity
\begin{gather}
  \frac{2\hbar}{m} y = \frac{\dot{\hat z}}{\hat z} -
     \frac{1}{\hat z ( \hat u + i\hat z )} = \frac{\dot\lambda}{\lambda} ,
\label{sm17}
\end{gather}
where $\hat u$ and $\hat z$ have been combined to form the complex variable
\begin{gather}
 \lambda = \hat u + i\hat z  .
\label{sm18}
\end{gather}
The quantity $\frac{2 \hbar}{m} y$ fulf\/ils the complex
nonlinear Riccati equation (here given for the harmonic oscillator)
\begin{gather}
 \frac{2\hbar}{m} \dot y + \left(\frac{2\hbar}{m}
            y\right)^2 + \omega^2 = 0,
\label{sm19}
\end{gather}
which can be linearized, using equation \eqref{sm17}, to a complex Newtonian equation
\begin{gather}
  \ddot{\lambda} + \omega^2 \lambda = 0
\label{sm20}
\end{gather}
for the variable $\lambda$ (for details see \cite{11,13}). The complex quantity $\lambda (t)$ can also be expressed
in polar coordinates as
\begin{gather}
  \lambda = \alpha  e^{i\varphi} = \alpha  \cos \varphi + i  \alpha  \sin \varphi,
\label{sm21}
\end{gather}
where relation \eqref{sm13} turns into
\begin{gather}
   \dot\varphi = \frac{1}{\alpha^2} ,
 \label{sm22}
\end{gather}
which has similarities with the conservation of angular momentum but, here,
for the motion of~$\lambda (t)$ in the {\it complex} plane.

Through equations \eqref{sm18}, \eqref{sm20} and \eqref{sm16} it is obvious that the imaginary part of
equation~\eqref{sm20}, up to a constant factor, is just the Newtonian equation of motion
for the wave packet maximum. The width of the wave packet \eqref{sm15}, or its
position uncertainty $\langle \tilde{x}^2\rangle  = \langle x^2\rangle  - \langle x\rangle ^2$, respectively, is
connected with the imaginary part of $\left(\frac{2\hbar}{m} y\right)$ via
$\frac{2\hbar}{m} y_I = \frac{\hbar}{2m \langle \tilde{x}^2\rangle}$. From $\lambda$ in
polar coordinates (see equation~\eqref{sm21}) and pursuing  equation~\eqref{sm17} and~\eqref{sm22} it follows that the variable of the complex Riccati equation \eqref{sm19} can be
written as
\begin{gather*}
\left(\frac{2\hbar}{m} y\right) = \frac{\dot{\lambda}}{\lambda} =
\frac{\dot{\alpha}}{\alpha} + i \dot{\varphi} = \frac{\dot{\alpha}}{\alpha} +
i \frac{1}{\alpha^2} = \left(\frac{2\hbar}{m} y_R \right) + i \left(\frac{2\hbar}{m}
y_I \right),
\end{gather*}
with $\alpha^2 = 2m \langle\tilde{x}^2\rangle / \hbar$. The imaginary part of equation~\eqref{sm19} agrees with equations~\eqref{sm22} and~\eqref{sm13}. The real
part of equation~\eqref{sm19} with the above-mentioned results leads, however, to a real
nonlinear dif\/ferential equation for the position uncertainty $\alpha (t)$,
\begin{gather}
\ddot\alpha + \omega^2(t)\alpha = \frac{1}{\alpha^3} .
\label{sm24}
\end{gather}

It is well-known in the literature \cite{14,15} that a pair of dif\/ferential
equations consisting of the Newtonian equation of motion \eqref{sm20}, but now only
for $\eta (t)$, and the nonlinear equation \eqref{sm24}, possesses a dynamical
invariant, the so-called Ermakov invariant,
\begin{gather}
I_L = \frac{1}{2} \left[ \left( \dot{\eta} \alpha - \eta \dot{\alpha}
  \right)^2 + \left( \frac{\eta}{\alpha} \right)^2 \right]  =
\textrm{const} = \frac12 \left(\frac{\alpha_0p_0}{m}\right)^2 ,
\label{sm25}
\end{gather}
that connects the classical position and momentum (or velocity) and their
uncertainties. This becomes even more obvious if one realizes that, with the help of $\alpha$, $\varphi$ and $\lambda$ and their time-derivatives (denoted by
overdots), the quantum mechanical uncertainties can now be expressed in the
following forms
\begin{gather}
\langle \tilde x^2\rangle   = \frac{\hbar}{2m}
        \alpha^2 = \frac{\hbar}{2m} \lambda \lambda^*,
\label{sm26}
\\
\langle \tilde p^2\rangle  =
         \frac{\hbar m}{2} (\dot \alpha^2 + \alpha^2
         \dot\varphi^2) = \frac{\hbar m}{2} (\dot\lambda \dot\lambda^*),
\label{sm27}
\\
\langle [\tilde x, \tilde p]_+\rangle  = \langle \tilde x \tilde  p + \tilde p \tilde x\rangle
        = \hbar \dot \alpha
        \alpha = \frac{\hbar}{2} \frac{\partial}{\partial t}
        (\lambda \lambda^*) ,
\label{sm28}
\end{gather}
where $\tilde{x} = x - \langle x\rangle$, $\tilde{p} = p - \langle p\rangle$.

In order to compare the time-dependent kernel \eqref{sm12} with the kernel $K(x,x')$ in
\eqref{sm08}, one must take into account that  $K(x,x')$ has been obtained via equations~\eqref{sm06} and~\eqref{sm07} which describe the transformation of $x$ and $p$ into the initial values
$x'$ and $p'$, whereas, $K(x,x',t,t')$ in \eqref{sm12} describes the inverse transformation from
$x'$ to $x$. This is expressed, e.g., by the dif\/ferent signs in the exponents of
\eqref{sm08} and \eqref{sm12}. In order to make a direct comparison, one must therefore take
the inverse transformation of \eqref{sm12}, obtained by changing the
sign in the exponent, and interchan\-ging~$\dot{\hat{z}}$ and~$\hat{u}$. Inserting this kernel into
equations~\eqref{sm06} and~\eqref{sm07}, one obtains the corresponding equations for the
time-dependent problem,
\begin{gather}
\dot{\hat{z}} x - \hat{z} \frac{p}{m} = \frac{x'}{\alpha_0},
\label{sm29}
\\
-\dot{\hat{u}} x + \hat{u} \frac{p}{m} =  - \frac{\alpha_0 p'}{m}
\label{sm30}
\end{gather}
or, in matrix notation,
\begin{gather}
\left(\begin{array}{c} \frac{x'}{\alpha_0}  \vspace{1mm}\\  - \frac{\alpha_0 p'}{m} \end{array}\right)  = \left(\begin{array}{cc} \dot{\hat{z}} & - \hat{z} \\ -\dot{\hat{u}} &
      \hat{u} \end{array}\right)   \left(\begin{array}{c} x \\ \frac{p}{m}
    \end{array}\right)  =  \mathbf{M}   \left(\begin{array}{c} x \\ \frac{p}{m}
    \end{array}\right).
\label{sm31}
\end{gather}

Again, the transformation matrix $\mathbf{M}$ has the determinant $\dot{\hat{z}} \hat{u} -
\dot{\hat{u}}  \hat{z} = 1$, which
corresponds, according to \eqref{sm22}, to a kind of conservation of angular momentum
(in a complex plane). However, dif\/ferent from the time-independent case, the
initial state is not only characterized by the initial position $x'$ and
momentum $p'$ but, also, by the corresponding initial uncertainties since
$\alpha_0 = (\frac{2m}{\hbar} \langle \tilde{x}^2\rangle _0)^{1/2}$ is proportional to the
initial position uncertainty and (for a minimum
uncertainty wave packet with $\langle \tilde{x}\rangle ^2_0 \langle \tilde{p}^2\rangle _0 = \hbar ^2 /4$)
the inverse $\frac{1}{\alpha_0} =  (\frac{2}{m \hbar} \langle \tilde{p}^2\rangle_0)^{1/2}$
is proportional to the initial momentum uncertainty, i.e.,
$\frac{x'}{\alpha_0} \propto \frac{x'}{\sqrt{ \langle \tilde{x}^2\rangle_0}}$, $\frac{\alpha_0 p'}{m} \propto \frac{p'}{\sqrt{\langle \tilde{p}^2\rangle_0}}$.

Following the procedure for the inverted propagator \eqref{sm12} outlined in
\cite{7}, by applying the Wigner transformation~\eqref{sm09},
one arrives at a similar result. The kernel $K(x,x',p,p',t,t')$ (being
time-dependent via $\hat{z} (t)$ and $\hat{u} (t)$), which provides the Wigner function $W(x,p,t)$ via
\begin{gather*}
 W(x,p,t) = \int^{+\infty}_{-\infty} dx'dp'\, K (x,x',p,p',t,t') W(x',p',t'=0)
\end{gather*}
with
\begin{gather}
  W(x',p',t'=0)  =  \frac{1}{\pi\hbar}\exp \left\{
                    -\frac{x'^2}{2 \langle\tilde{x}^2\rangle_0}  -\frac{p'^2}{2
                      \langle \tilde{p}^2\rangle _0} \right\} \nonumber\\
\phantom{W(x',p',t'=0)}{} =  \frac{1}{\pi \hbar}  \exp \left\{
                    -\frac{m}{\hbar} \left[ \left(\frac{x'}{\alpha_0}
                    \right)^2 +  \left(\frac{\alpha_0 p'}{m}\right)^2 \right]\right\} ,
\label{sm33}
\end{gather}
(see, e.g., \cite{11}), is given by the product of two delta
functions where now, however, $x'$ is replaced by $\frac{x'}{\alpha_0}$ and $p'$ is replaced by
$\frac{\alpha_0 p'}{m}$ and the transformed variables in the delta functions are determined by \eqref{sm29}, \eqref{sm30}, i.e.,
\begin{gather*}
 K (x, x',p,p',t,t'=0) = \delta \left[\left(\frac{x'}{\alpha_0} \right) -\left(\dot{\hat{z}} x -
                    \hat{z}\frac{p}{m}\right)\right] \delta \left[ \left(\frac{\alpha_0 p'}{m}\right) - \left(\hat{u}\frac{p}{m} - \dot{\hat{u}} x\right)\right].
\end{gather*}

Applying this kernel to the initial Wigner distribution function \eqref{sm33} yields
the function $W(x,p,t)$ as
\begin{gather*}
 W(x,p,t) = \frac{1}{\pi \hbar} \exp \left\{ -\frac{m}{\hbar} \left[ \left(\dot{\hat{z}} x -
                    \hat{z} \frac{p}{m}\right)^2 +
                    \left(\hat{u} \frac{p}{m} - \dot{\hat{u}} x\right)^2\right]
                    \right\}.
\end{gather*}

Using the def\/initions \eqref{sm18} and \eqref{sm21} of $\lambda$ and its relation to the uncertainties as
given in equations~\eqref{sm26}--\eqref{sm28}, f\/inally allows one to write
\begin{gather}
   W(x,p,t) = \frac{1}{\pi \hbar} \exp \left\{
                    -\frac{2}{\hbar^2} \left[ \langle \tilde p^2\rangle  x^2 -
                    \langle [\tilde x, \tilde p]_+\rangle   x  p  +
                    \langle \tilde x^2\rangle  p^2\right]\right\},
\label{sm36}
\end{gather}
where the time-dependence of the uncertainties is determined totally by the
time-dependence of $\hat{z} (t)$ and $\hat{u} (t)$. In the case of time-dependent Gaussian wave
packets, the classical time-dependence is expressed by the fact that the
maximum of the wave packet follows the classical trajectory. This is taken
into account by shifting the variables of position and momentum from~$x$ to
$\tilde{x} = x - \langle x\rangle $ and $p$ to $\tilde{p} = p - \langle p\rangle $. Since $\langle x\rangle $ and $\langle p\rangle$
are purely time-dependent quantities,
$\tilde{x}$~and~$\tilde{p}$ can
replace~$x$ and~$p$ in equations~\eqref{sm06},~\eqref{sm07} since these equations only contain
derivatives with respect to space, not time. So,~$x$ and~$p$ in \eqref{sm36} would
be replaced by~$\tilde{x}$ and~$\tilde{p}$ which would lead to the result obtained in \cite{16}
showing the connection between the exponent of the time-dependent Wigner
function and the dynamical Ermakov invariant that is connected with the parameters~$\hat{z}$ and~$\hat{u}$ of the time-dependent kernel $K(x,x',t,t')$ and has
been def\/ined in equation~\eqref{sm25} (for details see also~\cite{11}).

In the quantum mechanical phase space picture according to Wigner, this
results not only in changing initial position- and momentum-uncertainties into
their values at time~$t$ but, also, an additional contribution occurs from the
time-change of $\langle \tilde{x}^2\rangle $, or $\alpha ^2$, respectively, expressed by the term
proportional to $\langle [\tilde x, \tilde p]_+\rangle$, or $\dot{\alpha} \alpha$, respectively, in the exponent of $W(x,p,t)$.

All these quantum dynamical aspects are contained in the time-dependent
parameters $\hat{z}$ and $\hat{u}$, entering the transformation matrix in \eqref{sm31}. In
particular, the change of the position uncertainty (proportional to $\alpha$) is
taken into account by the parameter $\hat{u}$, which can be expressed as \cite{11}
\begin{gather}
  \hat u = \dot{\hat z} \alpha^2 - \hat z \dot \alpha
            \alpha = \left(\frac{m}{\alpha_0p_0}\right) [ \dot{\eta} \alpha^2 - \eta \dot \alpha
            \alpha ].
\label{sm37}
\end{gather}

For constant uncertainty $\alpha = \alpha_0$, $\hat{u}$ is simply proportional
to the classical velocity $\dot{\eta} (t)$, for $\dot{\alpha} \neq 0$,
however, the situation can become quite dif\/ferent. As an example the free
motion shall be discussed brief\/ly, since there $\alpha = \alpha (t)$, which is
expressed in the spreading of the corresponding wave packet solution. For this
purpose, also $\dot{\hat{u}}$ is now given in terms of $\eta$ and
$\dot{\eta}$ where the equations of motion \eqref{sm20} (for $\eta (t)$) and \eqref{sm24}
(for $\alpha (t)$) are applied. So it follows from \eqref{sm37} that
\begin{gather*}
  \dot{\hat{u}}  = \bigg(\frac{m}{\alpha_0p_0}\bigg) \left[ \dot{\eta} \dot{\alpha}
  \alpha - \eta \left( \dot{\alpha}^2 + \frac{1}{\alpha^2} \right) \right].
\end{gather*}

For constant $\alpha = \alpha_0$, all terms proportional to $\dot{\alpha}$
vanish and the transformation matrix in~\eqref{sm31} can be written as
\begin{gather*}
 \mathbf{M}   = \left(\begin{array}{cc}  \dot{\hat{z}} & - \hat{z} \\ -\dot{\hat{u}}
      & \hat{u} \end{array}\right)  = \left(\frac{m}{\alpha_0 p_0}\right) \left(\begin{array}{cc}  \dot{\eta} & - \eta \vspace{1mm}\\ \frac{1}{\alpha_0^2}
      & \alpha_0^2 \dot{\eta} \end{array}\right).
\end{gather*}

For the harmonic oscillator with constant width (note: there is another
solution with oscillating width) $\alpha = \alpha_0 = \frac{1}{\sqrt{\omega}}$
and $\eta (t) = \frac{v_0}{\omega} \sin \omega t$ ($\eta (0) = 0$), $\dot{\eta}
= v_0 \cos \omega t$ ($\dot{\eta} (0) = v_0 = \frac{p_0}{m}$),  $\mathbf{M}$~turns into
\begin{gather*}
 \mathbf{M}_{\rm HO} =\left(\frac{m}{\alpha_0 p_0} \right)\left(\begin{array}{cc}
      v_0 \cos \omega t  & -  \frac{v_0}{\omega} \sin \omega t \vspace{1mm} \\ v_0 \sin
     \omega t
      & \frac{v_0}{\omega} \cos \omega t \end{array}\right) = \left(\begin{array}{cc}
      \frac{1}{\alpha_0} \cos \omega t & -  \alpha_0 \sin \omega t \vspace{1mm}\\ \frac{1}{\alpha_0}  \sin
     \omega t
      &  \alpha_0  \cos \omega t \end{array}\right),
\end{gather*}
i.e., (up to the constant $\alpha_0$ that also occurs in the column vectors)
just the classical result is reproduced.

However, for the free motion with $\eta (t) = v_0 t$, $\dot{\eta} (t) = v_0$,
for constant $\alpha = \alpha_0$, one would obtain
\begin{gather*}
 \tilde{\mathbf{M}}_{\rm fr}  = \left(\frac{m}{\alpha_0 p_0} \right)\left(\begin{array}{cc}
      v_0  & -  v_0 t \vspace{1mm}\\ \frac{1}{\alpha_0^2} v_0 t
      & \alpha_0^2 v_0 \end{array}\right) ,
\end{gather*}
that is dif\/ferent from the classical situation where the matrix element in
the f\/irst column and second row would be zero. The consequence for a free motion
wave packet with constant width would be that the transformation matrix would
no longer describe a canonical transformation, since its determinant would no
longer be equal to 1 but
\begin{gather*}
\det ( \tilde{\mathbf{M}}_{\rm fr})  =  \left[ 1 + \left( \frac{t}{\alpha_0^2}
  \right)^2 \right]  ,
\end{gather*}
which just describes the time-dependence of the wave packet spreading.

For $\dot{\alpha} \neq 0$, the well-known time-dependence of the wave packet
width, given by $\alpha ^2 (t) = \alpha_0^2 \left[ 1 + \left( \frac{t}{\alpha_0^2}
  \right)^2 \right]$ and obtained as a solution of equation~\eqref{sm24} for $\omega = 0$, leads
to the correct transformation matrix
\begin{gather}
\mathbf{M}_{\rm fr}  = \left(\frac{m}{\alpha_0 p_0} \right)\left(\begin{array}{cc}
      v_0  & -  v_0 t \\  0
      & \alpha_0^2 v_0 \end{array}\right) ,
\label{sm43}
\end{gather}
with $\det (\mathbf{M}_{\rm fr}) = 1$. This shows explicitly the inf\/luence of the
time-dependence of the uncertainty~$\alpha$ on the transformation describing
the dynamics of the system.

It should also be mentioned that the determinant of $\mathbf{M}$, written in
terms of $\eta$, $\dot{\eta}$, $\alpha$ and $\dot{\alpha}$ takes just the form of
the Ermakov invariant, i.e.,
\begin{gather}
\mathbf{M}  = \left(\frac{m}{\alpha_0 p_0} \right)\left(\begin{array}{cc} \dot{\eta}  & - \eta \vspace{1mm}\\- \dot{\eta} \dot{\alpha} \alpha + \eta \left( \dot{\alpha}^2 + \frac{1}{\alpha^2} \right) & \dot{\eta} \alpha^2 - \eta \dot \alpha
            \alpha \end{array}\right)
\label{sm44}
\end{gather}
yields
\begin{gather}
\det (\mathbf{M})  = \left(\frac{m}{\alpha_0 p_0} \right)^2 \left[
\dot{\eta}^2 \alpha^2 - 2 \eta  \dot{\eta}  \alpha \dot{\alpha} + \eta^2 \left(
  \dot{\alpha}^2 + \frac{1}{\alpha^2} \right) \right] \nonumber\\
\phantom{\det (\mathbf{M})}{} = \left(\frac{m}{\alpha_0 p_0} \right)^2 \left[ (\dot{\eta} \alpha -
  \dot{\alpha} \eta )^2 + \left( \frac{\eta}{\alpha} \right)^2 \right]  .
\label{sm45}
\end{gather}

At this point it appears appropriate to discuss some dif\/ferences characteristic
of our method in comparison with other approaches that try to f\/ind a quantum analogue
to the classical canonical transformations in the case of time-dependent
quantum systems, particularly, when the Wigner phase space formulation is
applied.

So in \cite{17,18} the Wigner function for the free motion and Gaussian-shaped
distribution function is given by ($\hbar = 1$)
\begin{gather*}
 W(x,p,t) = \frac{1}{\pi } \exp \left\{ - \left[ \frac{1}{b} \left(x - \frac{p}{m} t
 \right)^2 + b p^2 \right] \right\}
\end{gather*}
with constant parameter $b$ (that would, in our notation, correspond to $b =
\frac{2}{\hbar^2} \langle \tilde{x}^2\rangle_0 = \frac{\alpha_0^2}{m \hbar} = \frac{\hbar^2}{2 \langle \tilde{p}^2\rangle_0}$
(for a minimum uncertainty wave packet)). The equivalence to the conserved
volume element in classical phase space is intended to be established by the
statement: ``this distribution is concentrated within the region where the
exponent is less than $1$ in magnitude'' and, for the choice of a coordinate
system where $b=1$, ``the above phase space distribution function is a circle
at $t=0$. As time progresses, the circle becomes a tilted ellipse while
preserving its area.'' This elliptic deformation is a canonical transformation
that corresponds to our transformation~\eqref{sm43} but, in the case of these authors'
description, the time-dependence in their transformation only originates from
the classical dynamics, transforming the initial position $x$ into $x -
\frac{p}{m} t$
(with constant~$p$). The correct spreading Gaussian wave packet for the
free motion, however, obviously has a time-dependent width, corresponding to a
time-dependent parameter $b$ in the notation of the references quoted. This
still holds if one considers the corresponding Wigner function. However, if
the width in position space is time-dependent, in our notation $\dot{\alpha}
\neq 0$, a
third term in the exponent of the Wigner function, namely, the one taking into
account the position-momentum-correlations (see equations~\eqref{sm28} and~\eqref{sm36}), must
occur. This term is missing in the above-mentioned approach.

In another approach \cite{19}, a relation between the Wigner function for
two-photon coherent states and the one for Glauber coherent states has been
established that looks quite similar to our transformations~\eqref{sm01} or~\eqref{sm31}. However, this method is based on combinations of creation and
annihilation operators. It has been shown recently \cite{20} that the Ermakov
invariant (that is, up to a constant factor, equivalent to the determinant of
our transformation matrix~\eqref{sm31}; see also~\eqref{sm45}), can be factorized into two
terms that are a kind of (complex) generalization of the creation and
annihilation operators and turn into these for $\dot{\alpha} = 0$ (for further details, see~\cite{20}). Therefore, also in the approach \cite{19} based on the usual
creation and annihilation operators, the time-dependence of the uncertainties
expressed by $\dot{\alpha} \neq  0$ is not taken into account.

It is also known in the literature that quantum uncertainties are related with
classical error margins and that, particularly for quadratic Hamiltonians, the
classical error margins satisfy the classical Hamiltonian equations
\cite{21}. Furthermore, in our case, it can even be shown that the quantum uncertainties obey equations of motion that can be derived from a
quantum-uncertainty Hamiltonian function that now provides the time-evolution
of the quantum uncertainties in a~canonical formalism. This Hamiltonian is
nothing but the ground state energy or, respectively, the energy
contribution of the momentum and position f\/luctuations to the overall energy
of the Gaussian wave packet solution for the system. To make these formal
aspects obvious, we write the energy of the system, calculated as the mean value
of the Hamiltonian operator using the Gaussian wave packet solutions of the
corresponding time-dependent Schr\"odinger equation, in the form (for the harmonic oscillator)
\begin{gather*}
\langle H_{\rm op}\rangle  = \frac{1}{2m}\langle p^2\rangle  + \frac{m}{2} \omega^2 \langle x^2\rangle  = \left(\frac{1}{2m}\langle p\rangle ^2 + \frac{m}{2} \omega^2 \langle x\rangle^2\right) + \left(\frac{1}{2m}\langle \tilde{p}^{2}\rangle  + \frac{m}{2} \omega^2 \langle \tilde{x}^{2}\rangle \right) \nonumber\\
\phantom{\langle H_{\rm op}\rangle}{} = \left(\frac{m}{2} \dot{\eta}^2 + \frac{m}{2} \omega^2 \eta^2 \right) + \left[\frac{\hbar}{4}\left(\dot\alpha^2 + \alpha^ 2  \dot{\varphi}^2 \right) + \frac{\hbar}{4} \omega^2 \alpha^ 2\right] \nonumber\\
\phantom{\langle H_{\rm op}\rangle}{} =E_{\rm cl} + \tilde{E} = (T_{\rm cl} + V_{\rm cl}) + (\tilde{T} + \tilde{V}).
\end{gather*}

In order to establish a Lagrangian/Hamiltonian formalism for the
uncertainties, we assume that a corresponding Lagrangian $\tilde{\cal L}$ can be written
as the dif\/ference between kinetic and potential energy f\/luctuations, but now
expressed in terms of the variables $\alpha$, $\varphi$ and the corresponding velocities
$\dot{\alpha}$, $\dot{\varphi}$, i.e.
\begin{gather*}
\tilde{\cal L} (\alpha, \dot{\alpha}, \varphi, \dot{\varphi}) = \tilde{T} - \tilde{V} =  \frac{\hbar}{4}\left(\dot\alpha^2 + \alpha^ 2
  \dot{\varphi}^2 - \omega^2 \alpha^ 2 \right) .
\end{gather*}

The corresponding Euler--Lagrange equations are then
\begin{gather*}
\frac{d}{dt} \frac{\partial \tilde{\cal L}}{\partial \dot{\varphi}} - \frac{\partial \tilde{\cal L}}{\partial \varphi} = 0,
\qquad
\frac{d}{dt} \frac{\partial \tilde{\cal L}}{\partial \dot{\alpha}} - \frac{\partial \tilde{\cal L}}{\partial \alpha} = 0 .
\end{gather*}
From the f\/irst equation follows $\frac{d}{dt} (\frac{\hbar}{2}
\alpha^2 \dot{\varphi}) =0 $, or, $\alpha^2 \dot{\varphi}  = {\rm const}$, in
agreement with equation~\eqref{sm22}; from
the second equation follows $\ddot\alpha + \omega^2 \alpha = \dot{\varphi}^2
\alpha  =  \frac{\rm const}{\alpha^3}$, equivalent  to equation~\eqref{sm24} (for ${\rm const}=1$).

The corresponding canonical  momenta are then given by
\begin{gather*}
\frac{\partial \tilde{\cal L}}{\partial \dot{\varphi}} = \frac{\hbar}{2}
\alpha^2 \dot{\varphi} = p_{\varphi},
\qquad
\frac{\partial \tilde{\cal L}}{\partial \dot{\alpha}} = \frac{\hbar}{2} \dot{\alpha} = p_{\alpha}.
\end{gather*}

With the help of these def\/initions, the quantum energy contribution $\tilde{E}
= \tilde{T} + \tilde{V}$ can
be written in a Hamiltonian form as
\begin{gather*}
\tilde{\cal H} (\alpha, p_{\alpha}, \varphi,
p_{\varphi}) = \frac{p_{\alpha}^2}{\hbar} + \frac{p_{\varphi}^2}{\hbar \alpha^2} +
\frac{\hbar}{4} \omega^2 \alpha^ 2 .
\end{gather*}

It is straightforward to show \cite{13} that the corresponding Hamiltonian
equations of motion again reproduce the results \eqref{sm22} and \eqref{sm24}, only now
expressed with the help of the canonical momenta. An interesting point is that
because of relation~\eqref{sm22}, i.e.~$\dot{\varphi} = \frac{1}{\alpha^2}$, the
canonical ``angular momentum'' $p_{\varphi}$ has the constant value
$p_{\varphi} = \frac{\hbar}{2}$, a value that does not usually describe an orbital
angular momentum but the non-classical angular momentum-type quantity spin.

Finally, using these results, the uncertainty product can be expressed as
\begin{gather*}
U =  \langle \tilde{x}^{2}\rangle \langle \tilde{p}^{2}\rangle = p_{\varphi}^2 + (\alpha p_{\alpha})^2.
\end{gather*}

In this context it should be mentioned that some authors (e.g., \cite{22, 23,  24}) assume that the role of the phase space volume in quantum mechanics is
played by the square root of the so-called ``invariant uncertainty product''
$\langle \tilde{x}^{2}\rangle \langle \tilde{p}^{2}\rangle  - \frac{1}{4} \langle [\tilde x, \tilde
p]_+\rangle ^2 = \frac{\hbar^2}{4}$, which is def\/initely a constant of motion since it is equivalent to our
quantity $p_{\varphi}^2$. So, this square root would just be $p_{\varphi} = \frac{\hbar}{2}
\alpha^2 \dot{\varphi}$ which, due to the
equivalence between equations~\eqref{sm13} and \eqref{sm22}, leads back to the requirement for the
determinant of our transformation matrix~\eqref{sm31} to be equal to~$1$.

\section{Conclusions and perspectives}\label{sec4}

In classical mechanics, canonical transformations can be characterized by
certain properties, like the conservation of a volume element in phase space
under these transformations, corresponding to Liouville's theorem that can
easily be checked by calculating the Wronskian determinant of the
transformation that should be equal to $1$.

In quantum mechanics, this conservation law appears to have its
correspondence
 in the conservation of a so-called ``invariant uncertainty product'' that
holds for systems with time-dependent and time-independent quantum
uncertainties since any explicit time-dependence of the uncertainties is
compensated for by subtracting a term proportional to $ \langle [\tilde x, \tilde
p]_+\rangle^2$. The remaining
conserved quantity corresponds to the conservation of $p_{\varphi}$, the ``angular
momentum'' for the motion of $\lambda (t)$ in the complex plane. Due to the equivalence
of equations~\eqref{sm13} and~\eqref{sm22}, this conservation law is identical with the requirement
that the determinant of our transformation matrix~\eqref{sm31} for the time-dependent
quantum problem must be equal to $1$. From equations~\eqref{sm44} and~\eqref{sm45}, it f\/inally
follows that this requirement is identical with the existence of a dynamical
invariant for the system, the so-called Ermakov invariant.

Another major result of our analysis is that, in the time-dependent quantum
mechanical problem, the transformation~\eqref{sm31} corresponding to the classical
linear canonical transformation~\eqref{sm01} and its time-independent quantum
mechanical analogue~\eqref{sm06},~\eqref{sm07} does not only transform the $\it{initial}$
 {\it position} and {\it momentum} into its values at a later time but, also, does the same
{\it simultaneously} with the {\it corresponding uncertainties}! In
how far this is connected with the existence of a Lagrangian/Hamiltonian
formulation of the dynamics of the quantum uncertainties will be further investigated.

So far, the discussion of the time-dependent case included only systems where the potential is at most
quadratic in its variables. This might not be as
restrictive as it seems at f\/irst sight since one may sometimes perform
canonical transformations to reduce a given Hamiltonian to a~quadratic form~\cite{25} which has been shown explicitly by Sarlet for some polynomial
Hamiltonians. To what extent this method can also be applied in our case requires
more detailed studies.

\subsection*{Acknowledgments}

Both authors would like to express their gratitude to the Instituto de F\'{\i}sica,
UNAM, that made possible the visit of the f\/irst author to Mexico. One of the
authors (D.S.) would like to thank Dr.~Robert Berger for valuable and
stimulating discussions.

\pdfbookmark[1]{References}{ref}
\LastPageEnding

\end{document}